\documentclass{article}
\usepackage{epsfig, graphicx, amsmath}
\usepackage{dsfont}
\usepackage{amsthm}
\usepackage{xcolor}
\usepackage{bm}
\usepackage{lipsum}
\usepackage{mwe}

\usepackage[a4paper, total={6in, 9.5in}]{geometry}
\usepackage[utf8]{inputenc}
\usepackage{authblk}

\usepackage[noadjust]{cite} % 'sort' and 'compress' are enabled by default
\title{Programming nonreciprocity and harmonic beam steering via a digitally space-time-coded metamaterial antenna}

\author{Shaghayegh Vosoughitabar}
\author{Chung-Tse Michael Wu}

\affil{Department of Electrical and Computer Engineering, 
\\ Rutgers, the State University of New Jersey}

\date{}  
\begin{document}
\maketitle

\begin{abstract}
Recent advancement in digital coding metasurfaces incorporating spatial and temporal modulation has enabled simultaneous control of electromagnetic (EM) waves in both space and frequency domains by manipulating incident EM waves in a transmissive or reflective fashion, resulting in time-reversal asymmetry. Here we show in theory and experiment that a digitally space-time-coded metamaterial (MTM) antenna with spatiotemporal modulation at its unit cell level can be regarded as a radiating counterpart of such digital metasurface, which will enable nonreciprocal EM wave transmission and reception via surface-to-leaky-wave transformation and harmonic frequency generation. Operating in the fast wave (radiation) region, the space-time-coded MTM antenna is tailored in a way such that the propagation constant of each programmable unit cell embedded with varactor diodes can toggle between positive and negative phases, which is done through providing digital sequences by using a field-programmable gate array (FPGA). Owing to the time-varying coding sequence, harmonic frequencies are generated with different main beam directions. Furthermore, the space time modulation of the digitally coded MTM antenna allows for nonreciprocal transmission and reception of EM waves by breaking the time-reversal symmetry, which may enable many applications, such as simultaneous transmit and receive, unidirectional transmission, radar sensing, and multiple-input and multiple-output (MIMO) beamformer. 
\end{abstract}
\section{Introduction}

 Owing to the fact that the magnetic or electric properties can be modulated periodically in both spatial and temporal domain  \cite{Taravati2020}, several space-time modulated architectures have been proposed that can work as a mixer, circulator, transceiver, or nonreciprocal components  \cite{Taravati2020, Taravati2017, Qin2014, Alu2016, Taravati2022, Alu2016_2, Sha2021, Bukhari2019, Taravati2018, Ramaccia2014, Taravati2020_2}. Very recently, digitally-coded metasurfaces have appeared to be an effective device to control electromagnetic (EM) waves through manipulating the reflection or transmission phase of each programmable unit cell by switching the integrated pin diode between on ``$1$'' and off ``$0$'' states \cite{Lei2018,Ma2022,Ma2019}. By changing coding sequences periodically in the time domain, a space-time coding digital metasurface can be realized to provide interesting applications \cite{zhang2018,zhang2019_ninrecip,Dai2021,Zhang2022,Zhao2018,Dai2020,Dai2019,Li2022,Wu2020,Dai2020_2,Zhang2021,Liu_2021,Cui2014}. For example, in \cite{zhang2018} digitally coded space-time modulation is leveraged to control the pattern of reflected harmonic waves from the metasurface. In \cite{zhang2019_ninrecip} a space-time-coding digital metasurface is proposed to break the time reversal symetry by properly designing the digital coding sequences, thereby enabling nonreciprocal wave reflection and frequency conversion.

On the other hand, time modulation has also been applied to antenna arrays \cite{Maneiro2017}, in which a predetermined periodic sequence can achieve side lobe reduction for the fundamental frequency pattern \cite{Keen1997}. Moreover, generated harmonic components can be leveraged to provide physical layer security in wireless communication as well as enable radar detection \cite{He2018, SHAN2018,Euzi2014,Poli2011,zhuu2014,Lii2022,Sina2019,Nusenu2021,Ali2022,Tennant2007,Kanbaz2021,Gangli2010,Ali2023_2}. Nevertheless, for conventional time modulated antenna arrays, pin diodes are usually employed as RF switches that decrease the power efficiency due to power dissipation in their off state \cite{Yang2004}.

\begin{figure*}[tbhp]
\centering
%17.8_10.1
\includegraphics[width=13.6cm,height=7.7cm]{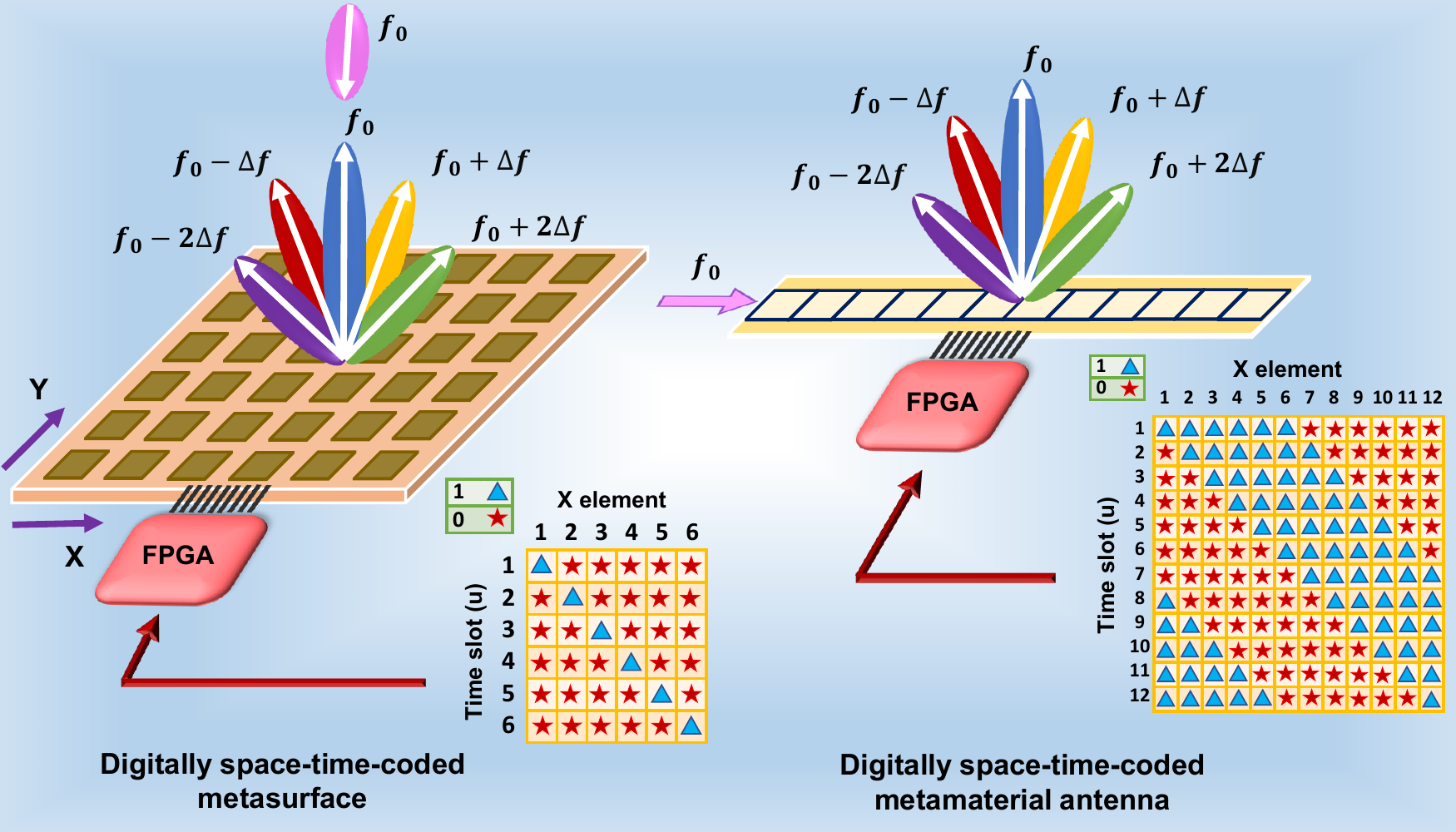}
\caption{Space-time-modulated metamaterial (ST-MTM) leaky wave antenna as a radiating counterpart of the space-time-modulated metasurface. The space time coding matrices are generated by an FPGA. For the metasurface, the coding sequence is uniform along the Y axis, where ``$0$'' and ``$1$'' refer to the reflection phase of $0$° and $180$°, respectively. For the proposed ST-MTM leaky wave antenna, ``$0$'' and ``$1$'' refer to a negative and positive phase constant ($\beta$), respectively.}\label{fig:dual}
%\end{SCfigure*}
\end{figure*}

In this work, we propose a programmable space-time digitally modulated composite right/left-handed (CRLH) metamaterial (MTM) leaky wave antenna \cite{itoh2005electromagnetic} that can be viewed as a radiating counterpart of the space-time-modulated digital metasurface as shown in Fig. \ref{fig:dual}. 
To illustrate, MTMs are artificial EM materials with novel effective medium properties that may not be available in nature. One type of MTM antenna is so-called CRLH transmission line leaky-wave antennas  exhibiting continuous backfire-to-endfire frequency-dependent beam scanning with a true broadside main beam. Due to the passive nature, conventional CRLH MTM antennas are reciprocal structures operating at the fundamental dominant mode, which cannot separate transmit and receive signals without an external duplexer. To this end, we show a space-time modulated CRLH MTM antenna with digitally programmable unit cells can enable nonreciprocal EM wave transmission and reception. Moreover, harmonic beam controlling can also be achieved through feeding appropriate coding sequences. 

%The new finding will pave the way to develop next-generation intelligent antennas.

In the case of a space-time-modulated metasurface, an incoming signal at $f_0$ frequency is illuminated to the surface, whereas the generated harmonics are reflected to the free space. By manipulating the phase and magnitude distribution of the surface through feeding appropriate periodic sequences to each unit cell, harmonic beamforming capability can be achieved. On the other hand, for our proposed space-time metamaterial (ST-MTM) antenna, a signal at $f_0$ is injected to the antenna input port and the generated harmonics will radiate to the free space. We show theoretically and experimentally that our proposed architecture enables harmonic scanning, simultaneous transmit and receive, and nonreciprocal behavior by providing proper spatiotemporal coding sequences for modulating the phase constant of each programmable CRLH MTM unit cell in a periodic fashion. %It is noted that in our proposed structure there is no need for any phase shifter or pin diode, owing to the fact that 
For proof-of-concept realization, varactors are incorporated into each CRLH unit cell to change its phase constant between the positive ``state 1'' and negative ``state 0'' values by feeding two specific bias voltages \cite{Lim2005}, in which the prespecified periodic sequences are generated by an FPGA and employed as varactors’ control voltages. To the best of the authors’ knowledge, this is the first digitally space-time-coded programmable CRLH MTM antenna exhibiting nonreciprocal and harmonic beamsteering capabilities, which can open up a new paradigm of next-generation intelligent antennas.

%\begin{SCfigure*}[\sidecaptionrelwidth][t]

%\begin{figure}[tbhp]
%\centering
%\includegraphics[width=1\linewidth]{Fig.1 Dual.pdf}
%\caption{Space-time-modulated leaky wave antenna as a duality of the space-time-modulated metasurface. The space time coding matrices are also shown. For the metasurface, the coding sequence is the same for all the elements along the Y axis. In the metasurface, ``$0$'' refers to the reflection phase $0$° and ``$1$'' refers to the reflection phase $180$°. In the leaky wave antenna, ``$0$'' refers to a negative phase constant ($\beta$) and ``$1$'' refers to the positive one.}
%\label{fig:dual}
%\end{figure}

%\begin{SCfigure*}[\sidecaptionrelwidth][t]
%\centering
%\includegraphics[width=11.4cm,height=11.4cm]{Fig0.pdf}
%\caption{This legend would be placed at the side of the figure, rather than below it.}\label{fig:side}
%\end{SCfigure*}
\section*{Results}

\subsection*{Formulation of the radiation pattern of a digitally coded space-time modulated metamaterial (ST-MTM) leaky wave antenna}
We first consider a programmable CRLH MTM unit cell consisting of interdigital capacitance and shunt-stub inductance with embedded  varactor diodes shown in Supplementary, Fig. \ref{fig:SI1}A. The absolute value of phase constant ($\beta$) of the designed unit cell multiplied by the unit cell length $p$ versus frequency, known as the dispersion curve, for two different bias voltages of the integrated varactors is illustrated in Supplementary, Fig. \ref{fig:SI1}B. The frequency region under the air line is regarded as the fast wave region in which the CRLH unit cells can radiate. By changing the varactor bias voltage, one can manipulate the dispersion curve of the MTM unit cell, where the phase constant can alternate its polarity under two bias voltages (Bias 1 and Bias 0) at a given frequency located in the fast wave region.

As such, both positive and negative phase constants can be achieved by simply changing the control voltage from Bias 1 to Bias 0, thereby realizing a binary digital MTM radiating unit cell. We denote that state ``$1$'' provides positive $\beta$ and state ``$0$'' provides negative $\beta$. A leaky wave antenna can thus be achieved by cascading such digital binary CRLH MTM cells, where the input signal operating in fast wave region radiates out as it propagates along the MTM structure. Moreover, by imposing the time modulation, the phase constant of each unit cell, i.e., $\beta$, changes with time in a periodic manner. By incorporating the array factor approach \cite{Yong2021}, in the transmit mode, the radiation pattern of a ST-MTM leaky wave antenna with digital coding can be expressed as:
\begin{align} 
\label{eq:RadiationPattern}
R(\theta,t) = \ S(t)\sum_{n = 1}^{N}{{I_{0}e^{- \alpha(n - 1)p}e}^{jk_{q}(n - 1)pcos\theta}U_{n}(t)}
\end{align}

\begin{align} 
\label{eq:Uswitch}
U_{n}(t) = e^{- j\sum_{m = 1}^{n}\varphi_{m(t)}} = \sum_{u = 1}^{L}{\gamma_{n}^{u}H_{u}(t)}
\end{align}

\begin{align} 
\label{eq:Hswitch}
H_{u}(t) = \begin{cases}
  1  & \frac{(u - 1)T}{L} \leq t \leq \frac{uT}{L}\\
  0 &  \text{ Otherwise}
\end{cases}
\end{align}

\begin{align} 
\label{eq:gammaswitch}
\gamma_{n}^{u} = \prod_{m = 1}^{n}e^{- j\varphi_{m}^{u}}, \varphi_{m}^{u} = \left\{\begin{matrix}
\beta_{0}p,\ \ \text{State}\ 0, \\ \beta_{1}p,\ \ \text{State}\ 1,\  \\
\end{matrix} \right. \ \text{in}\ \frac{(u - 1)T}{L} \leq t \leq \frac{T}{L}\ 
\end{align}

\begin{align} 
\label{eq:HFswitch}
H_{u}(t) = \sum_{q = - \infty}^{\infty}{a_{u}^{q}e^{\frac{j2\pi qt}{T}}}
\end{align}
where $S(t)$ is the input signal, which is a sinusoidal wave with frequency $f_0$. $N$ is the number of CRLH cells of the leaky wave antenna. $p$   is length of the unit cell, $\alpha$ is leakage factor, and $L$ is length of the sequence in one period ($T=1/\Delta f$). Moreover, $k_q=2\pi(f_0+q\Delta f)/ c$ %\textcolor{red}{(define c)}
is the propagation constant for 
%\textcolor{red}{(mw: use this notation for all following terms)} 
the $q^{th}$ radiated harmonic frequency, where $c$ is the light speed. Generated harmonics will radiate into the free space with different propagation constants when modulation frequency ($\Delta f$) is not much smaller than the input RF signal frequency ($f_0$). Each period is divided to $L$ time slots and the phase shift of the signal reaches to the $n^{th}$ unit cell, is a periodic function of time and over the $u^{th}$ time slot is defined as \eqref{eq:gammaswitch}. After expansion of $H_u(t)$ in the form of Fourier series shown in \eqref{eq:HFswitch} and substituting it in \eqref{eq:RadiationPattern} and doing some manipulations, finally we obtain (see Supplementary for detailed derivation):

\begin{align} 
\label{eq:R1}
R(\theta,t)=&\sum_{q = - \infty}^{\infty}e^{- j2\pi t(f_{0} - q\mathrm{\Delta}f)}sinc(\frac{\pi q}{L})e^{\frac{j\pi q}{L}}\sum_{u = 1}^{L}e^{\frac{- j2\pi qu}{L}}\sum_{n = 1}^{N}\frac{\gamma_{n}^{u}}{L}{I_{0}e^{- \alpha(n - 1)p}e}^{jk_{q}(n - 1)pcos\theta}
\end{align}

From \eqref{eq:R1}, it can be seen that the radiated patterns entail both radiation at the fundamental frequency ($f_0$) and generated harmonic frequencies ($f_0-q\Delta f$). Therefore, the far-field radiation pattern of the antenna at the $q^{th}$ generated harmonic frequency can be expressed as:

\begin{align} 
\label{eq:R2}
&R^{TX}_{q}(\theta) =sinc(\frac{\pi q}{L})e^{\frac{j\pi q}{L}}\sum_{u = 1}^{L}{e^{\frac{- j2\pi qu}{L}}\sum_{n = 1}^{N}\frac{\gamma_{n}^{u}}{L}}{I_{0}e^{- \alpha(n - 1)p}e}^{jk_{q}(n - 1)pcos\theta}
\end{align}

On the other hand, in the receive mode, assuming that the signal at $f_0$ is illuminated to the digital ST-MTM antenna from the $\theta$ angle, and generated harmonics are received from the same port of signal injection in the transmit mode, these patterns can be written as: (see Supplementary for detailed derivation):

\begin{align} 
\label{eq:R3}
&R^{RX}_{q}(\theta) =sinc(\frac{\pi q}{L})e^{\frac{j\pi q}{L}}\sum_{u = 1}^{L}{e^{\frac{- j2\pi qu}{L}}\sum_{n = 1}^{N}\frac{\gamma_{n}^{u}}{L}}{I_{0}e^{- \alpha(n - 1)p}e}^{jk_{0}(n - 1)pcos\theta}
\end{align}

The only difference of equation \eqref{eq:R3} with \eqref{eq:R2} is replacing $k_q$ with $k_0=2\pi f_0/\ c$ since the incident signal operates at $f_0$ in the free space and harmonics are generated when the signal is traveling inside the ST-MTM antenna. The pattern of the harmonic frequencies can be controlled by properly feeding the periodic sequence to each unit cell, thereby enabling several functionalities such as harmonic beam scanning, simultaneous transmit and receive, and nonreciprocity. The schematic of the proposed ST-MTM antenna consisting of programmable CRLH unit cells is illustrated in Fig. \ref{fig:HScanning}A.

%\begin{SCfigure*}[\sidecaptionrelwidth][t]

\subsection*{Programmable harmonic beam scanning} 
\begin{figure*}[bth]
\centering
%\includegraphics[width=17.8cm,height=19cm]{Fig2_HS.pdf}
%17.8_18.7
\includegraphics[width=13.6cm,height=14.2cm]{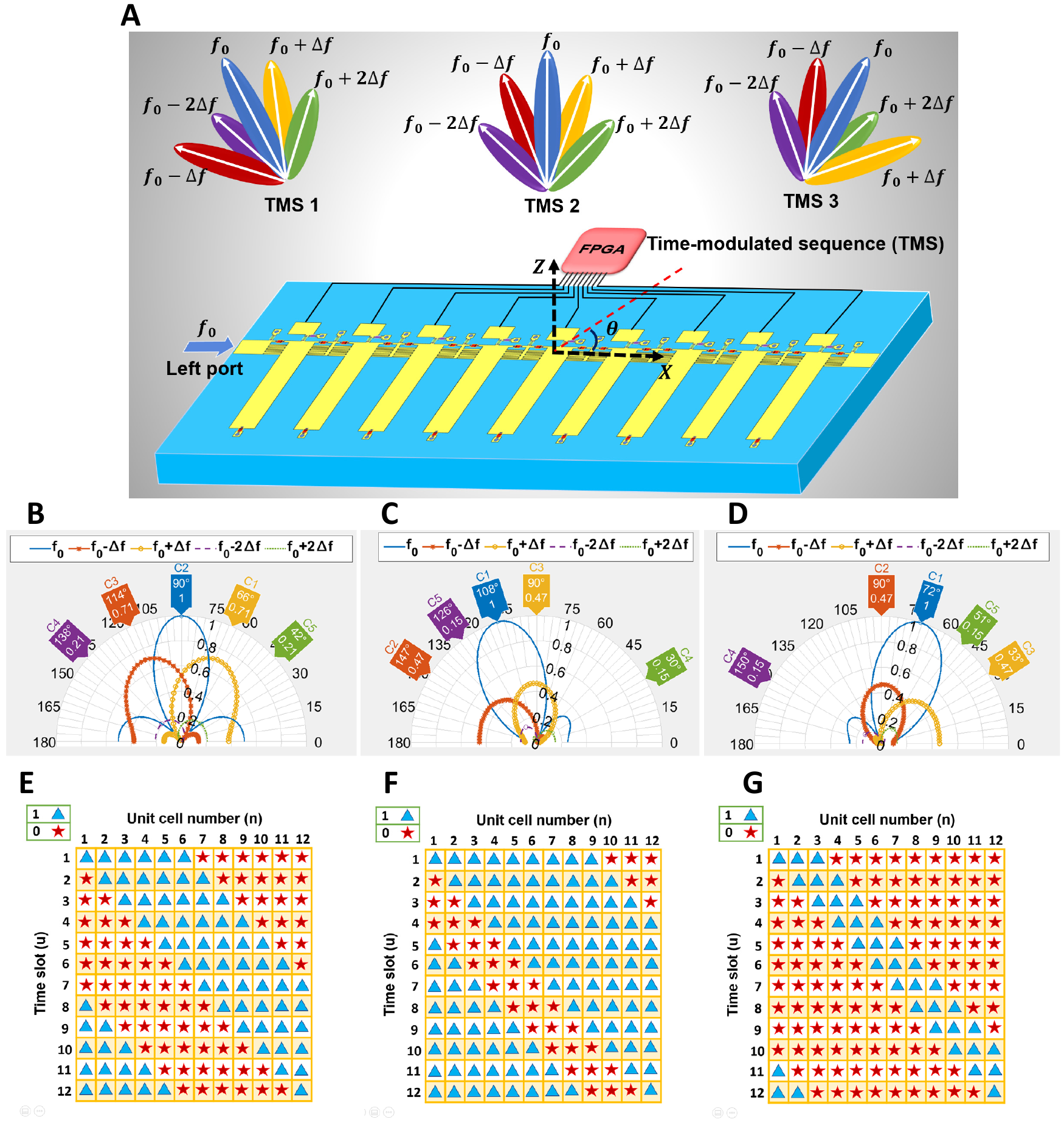}
\caption{(A) Programmable harmonic beam scanning of the proposed ST-MTM antenna with different space-time coding sequences. Simulated normalized harmonic patterns (dB) when the signal is injected from the left port. (B) for sequence 1111111000000. (C) for sequence 111111111000. (D) for sequence 111000000000. State of each unit cell in different time slots in one period for (E) figure B. (F) figure C. (G) figure D.}
\label{fig:HScanning}
\end{figure*}

As also shown in Fig. \ref{fig:HScanning}A, the proposed ST-MTM antenna can exhibit
dynamic harmonic beam steering, in which the fundamental and harmonic
radiation patterns are plotted in Fig. \ref{fig:HScanning}B-D when the signal at the
fundamental tone is injected from the left port with \emph{N=L}=12.
According to the simulated dispersion diagram of the designed programmable
CRLH unit cell in Bias 0 and Bias1 states shown in Supplementary, Fig. \ref{fig:SI1}B, the negative and positive phase shift values
(\(\beta_{0}p\ ,\beta_{1}p\)) around 2.05 GHz located in the fast wave
are 24 and -24 degrees, respectively. In Fig. \ref{fig:HScanning}B the sequence is
111111000000 for the 12 cells in the first time slot and then it is
circulated one bit backward for the next time slots as illustrated in
Fig. \ref{fig:HScanning}E to set the main beam direction of fundamental frequency in
broadside. As observed, harmonic beam scanning is achieved by using this sequence, i.e. each harmonic frequency component has its own main beam direction. By changing the sequence to 111111111000 and
111000000000, the main beam of the fundamental frequency and harmonic
components can be shifted to the right and left side, respectively, as shown in Fig. \ref{fig:HScanning}C and \ref{fig:HScanning}D. Fig. \ref{fig:HScanning}F and \ref{fig:HScanning}G 
%\textcolor{red}{(add reference)}
depict the space time coding
matrix corresponding to the patterns shown in Fig. \ref{fig:HScanning}C and \ref{fig:HScanning}D
%\textcolor{red}{(add reference)}. 
The resulting harmonic beam scanning can be utilized in various applications, including multipoint communication purposes or radar sensing systems for multi target detection.

%\end{SCfigure*}

\begin{figure*}[ht]
\centering
%17.8_19.3
%\includegraphics[width=17.8cm,height=19cm]{Fig.3_Final3.pdf}
\includegraphics[width=12.6cm,height=13.7cm]{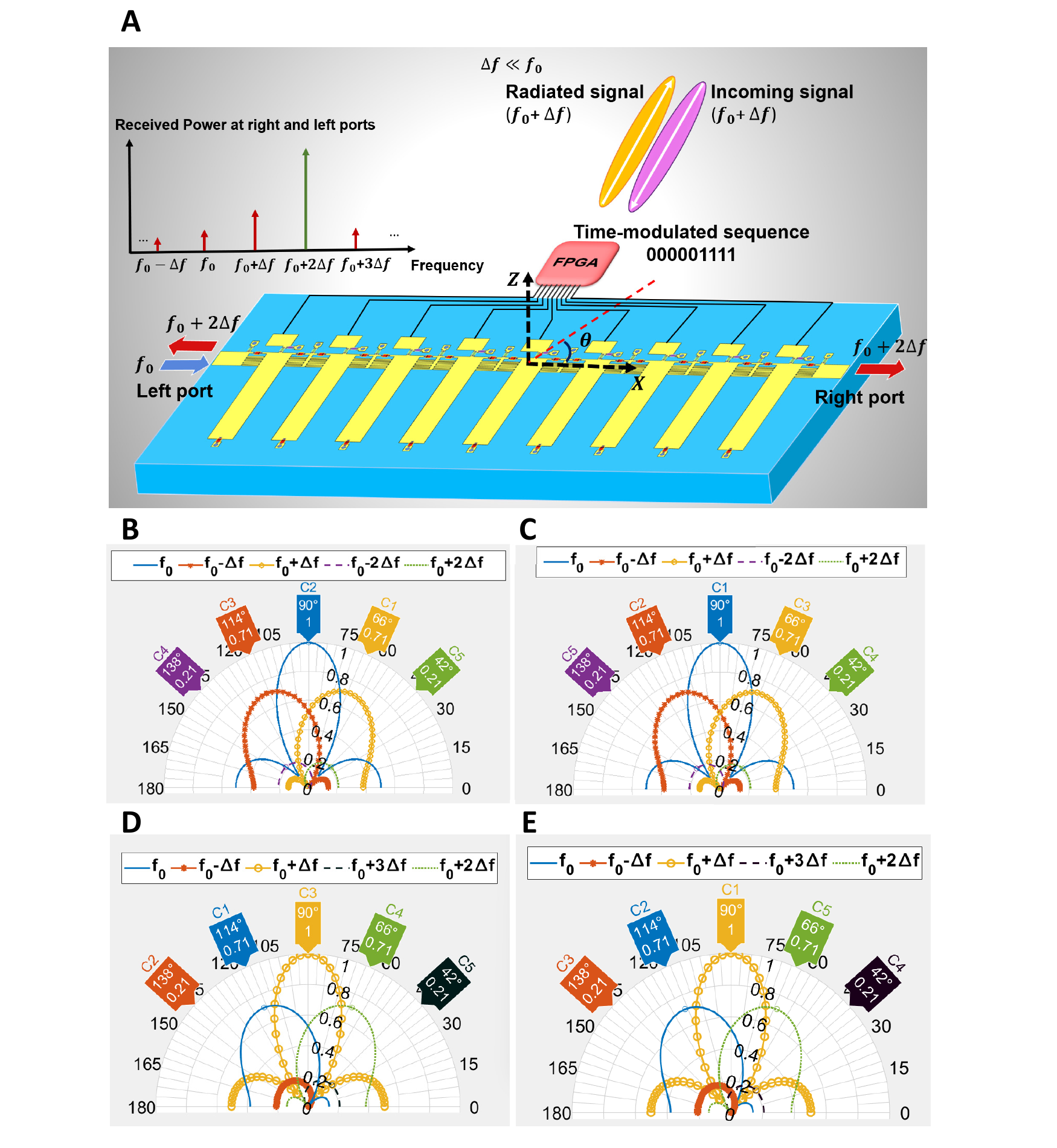}
\caption{(A) simultaneous transmit and receive by ST-MTM antenna when (\(\mathrm{\Delta}f \ll f_{0}\)). Simulated normalized harmonic patterns (dB) for sequence 111111000000. (B) when the signal is injected from the left port at $f_0$. (C) when the signal is injected from the right port   at $f_0$. (D) when the signal is illuminated at $f_0+\Delta f$ and harmonics are received from the left port. (E) when the signal is illuminated at $f_0+\Delta f$ and harmonics are received from the right port.
%\textcolor{red}{(change line colors for D and E to be consistent with BC in terms of harmonic frequencies.)}
}
\label{fig:Stransmit}
\end{figure*}

\subsection*{Simultaneous transmit and receive}
The proposed digitally coded ST-MTM antenna is essentially a 2-port structure and can transmit and receive the information simultaneously as depicted in Fig. \ref{fig:Stransmit}A. Fig. \ref{fig:Stransmit}B illustrates the case where the fundamental tone signal is
injected from the left port, which has the same patterns as Fig. \ref{fig:HScanning}B.

Fig. \ref{fig:Stransmit}C illustrates the patterns for injecting the signal from the right
port with the same space-time sequence used in Fig. \ref{fig:Stransmit}B. According to
\eqref{eq:R2} and \eqref{eq:R3}, it can be observed that if
\(\mathrm{\Delta}f \ll f_{0}\) for each harmonic frequency, the
far-field pattern when the signal with \(f_{0}\) frequency is injected
from the left/right port is the same as the received power of that
harmonic in different angles from the same port when the signal is
illuminated at \(f_{0}\) frequency to the ST-MTM antenna. Therefore, we
can consider Fig. \ref{fig:Stransmit}C patterns for illuminating the signal at \(f_{0}\)
frequency from different angles and receiving the fundamental signal and
harmonic components from the right port. As clearly observed in Fig. \ref{fig:Stransmit}B
and \ref{fig:Stransmit}C the patterns are the same for each harmonic frequency when the
signal with frequency \(f_{0}\) is injected from the left and when the
signal is illuminated to the ST-MTM antenna, and power levels are
recorded from the right port.

Considering this interesting property, the idea of simultaneous transmit
and receive can be made feasible as illustrated in Fig. \ref{fig:Stransmit}A. Assuming in
the transmit mode, the signal is injected at \(f_{0}\) frequency from
the left port, a transceiver located in the main beam direction of
\(f_{0} + \mathrm{\Delta}f\) pattern can receive the signal at
\(f_{0} + \mathrm{\Delta}f\) and transmit another signal with this
frequency at the same time. In this case under the assumption
\(\mathrm{\Delta}f \ll f_{0}\) , the generated normalized harmonic
patterns extracted from the left and right ports are shown in Fig. \ref{fig:Stransmit}D
and \ref{fig:Stransmit}E, respectively. According to these figures,
\(f_{0} + \mathrm{\Delta}f + \mathrm{\Delta}f\), which is the first
harmonic in the receive mode, will be the dominant harmonic frequency in
this direction and propagates to both right and left ports. As such, the
received information can be extracted from the left or right port at
\(f_{0} + 2\mathrm{\Delta}f\). %\textcolor{red}{(also from the left port?)}

%It is worth mentioning that in the transmit mode when the signal is injected from the left port at \(f_{0}\), the level of the signal of \(f_{0} + 2\mathrm{\Delta}f\) is negligible in the right port in comparison with the receive mode if the length of the leaky wave antenna is large enough. Therefore, good isolation with respect to the \(f_{0} + 2\mathrm{\Delta}f\) frequency is achievable between the transmit and receive mode. Moreover, in the receive mode, the power level of the downconverted signal at \(f_{0}\) reaching to the left port is much smaller than the input signal injected from the left port at \(f_{0}\), which is also of importance for the simultaneous transmit and receive functionality.

\subsection*{Nonreciprocal behavior of digitally coded ST-MTM antenna}
\begin{figure*}[tb]
\centering
%17.8_16.5
%\includegraphics[width=17.8cm,height=18.7cm]{Fig.4_1.pdf}
\includegraphics[width=13.3cm,height=12.4cm]{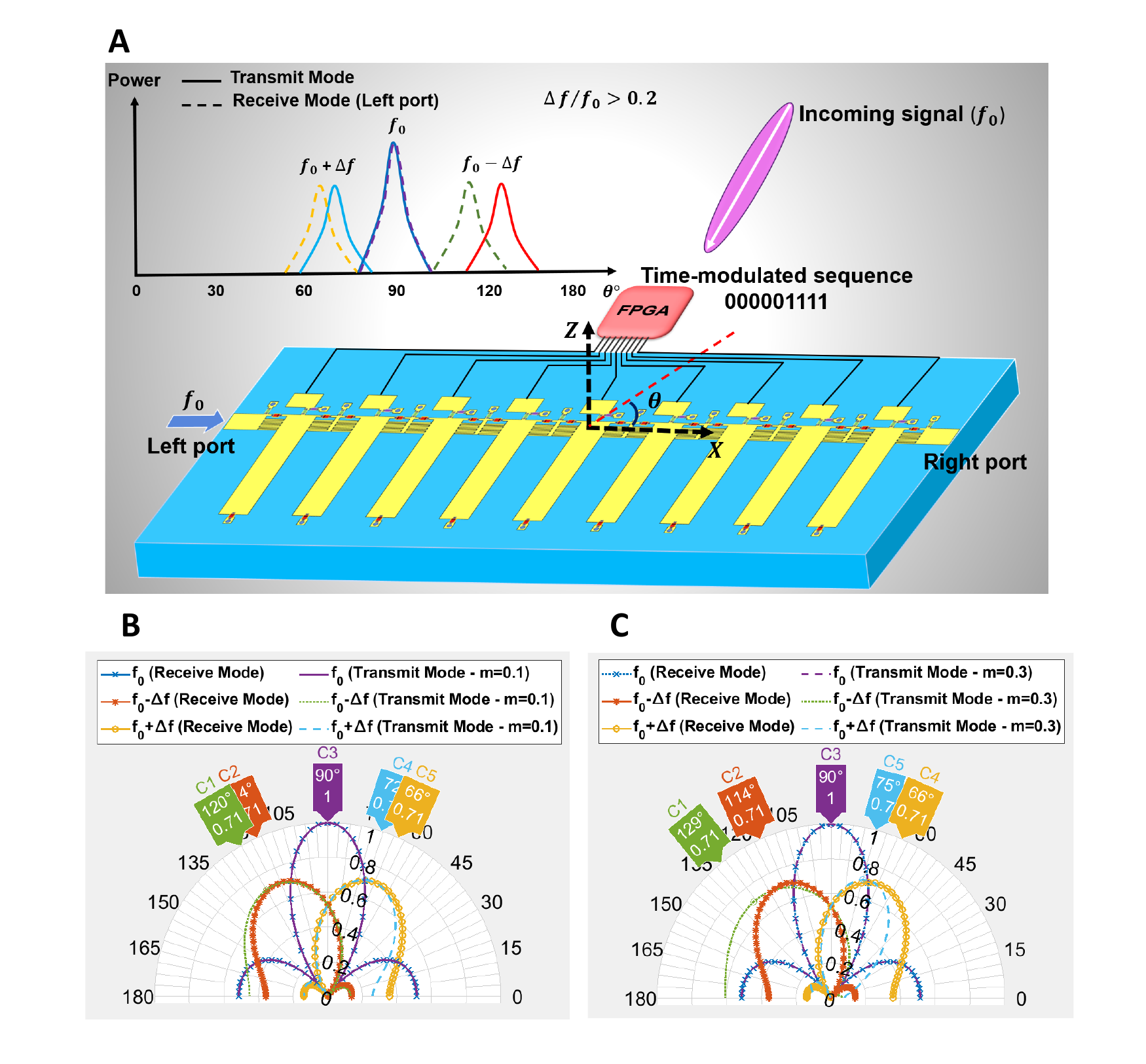}
\caption{(A) Nonreciprocal behavior for transmission and reception of the signal at harmonic frequencies when \(\mathrm{\Delta}f/f_{0} > 0.2\). Simulated normalized harmonic patterns in transmit mode and receive mode (dB). (B) for \(\mathrm{\Delta}f/f_{0} = 0.1\). (C) for \(\mathrm{\Delta}f/f_{0} = 0.3\). }
\label{fig:Nonreciprocity}
\end{figure*}

Based on \eqref{eq:R2} and \eqref{eq:R3}, when the modulation frequency
(\(\mathrm{\Delta}f\)) is not much smaller than the fundamental
frequency (\(f_{0}\)), the main beam direction of harmonic patterns in
the transmit mode will be different than their maximum power direction
extracted from the same port in the receive mode shown in Fig. \ref{fig:Nonreciprocity}A. Fig. \ref{fig:Nonreciprocity}B and \ref{fig:Nonreciprocity}C. illustrate the transmission and reception patterns for
different modulation frequencies. According to the results, increasing
\(\mathrm{\Delta}f/f_{0}\) leads to increasing the difference between
the main beam direction of transmission and reception patterns of each harmonic frequency. This property depicts the nonreciprocal behavior of
the ST-MTM antenna for \(m = \mathrm{\Delta}f/f_{0} > 0.2\).

Moreover, when \(f_{0}\) is injected from the left port in transmit
mode, consider the main beam direction of generated harmonic pattern
with \(f_{0} + \mathrm{\Delta}f\) frequency in
\(\theta_{1} = 66{^\circ}\) shown in Fig. \ref{fig:Stransmit}B. In the receive mode under
the assumption \(\mathrm{\Delta}f \ll f_{0}\), if a signal with
\(f_{0} + \mathrm{\Delta}f\) frequency is illuminated to the ST-MTM
antenna in different angles, the harmonic patterns extracted from left
port are like Fig. \ref{fig:Stransmit}D by considering \(f_{0} + \mathrm{\Delta}f\) as
fundamental frequency. As such, illuminating the signal in
\(f_{0} + \mathrm{\Delta}f\) frequency from \(\theta_{1}\) angle to the
antenna leads to reception of a very small power in \(f_{0}\) frequency
from the left port, thereby exhibiting nonreciprocal operation.

\subsection*{Experimental verification}
\begin{figure*}[tb]
\centering
\includegraphics[width=16cm,height=16.8cm]{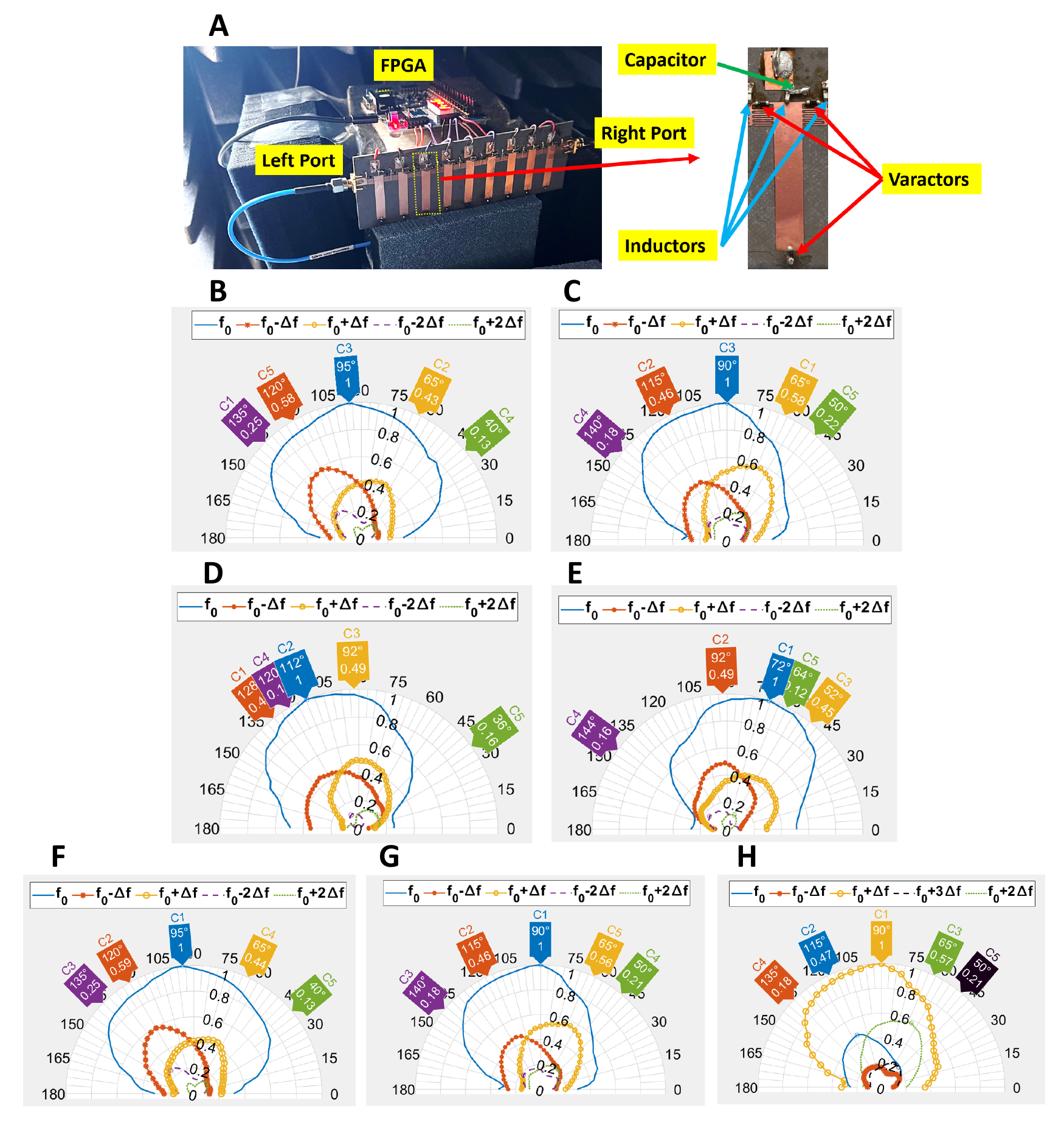}
\caption{
(A) Fabricated prototype of the proposed ST-MTM antenna consisting of 9 programmable CRLH unit cells with the close-up view of the MTM unit cell. (B) Measured normalized harmonic patterns (dB) in the transmit mode with sequence 111100000 when the signal is injected from the left port. (C) The signal is injected from the right port. (D) The signal is injected from the left port with sequence 111111100. (E) sequence 110000000. (F) Receive mode: The ST-MTM antenna is illuminated by a horn antenna at $f_0$ for sequence 111100000, and signals are received from the left port.(G) from the right port. (H) The ST-MTM antenna is illuminated by a horn antenna at $f_0+\Delta f$, and the signals are received from the right port for sequence 111100000. In all cases, $f_0$ is $2.05$ GHz and $\Delta f$ is $1.58$ MHz.
}
\label{fig:Measurement}
\end{figure*}

As illustrated in Fig. \ref{fig:Measurement}A, a prototype of the proposed digitally space-time-coded MTM antenna with 9 unit cells is fabricated to validate the aforementioned concepts. Each unit cell includes three varactors sharing a common bias voltage. According to the dispersion diagram of the unit cell around $2.05$ GHz the phase constant can toggle between positive and negative values by changing the bias of varactors from 1 to 0 state. The number of time slots (\emph{L}) are considered 9 (same as the number of unit cells), and the sequence in the first time slot is shifted by one bit for the next time slot by using an FPGA shown in Fig. \ref{fig:Measurement}A. The
modulation frequency we apply here is $1.58$ MHz, which is much smaller
than the fundamental frequency (\(\mathrm{\Delta}f \ll f_{0}\)).

Measured $S$-parameters of the sample are plotted in Supplementary, Fig. \ref{fig:SI2},
which depicts good S11 and S22 around $2$ GHz. In the transmit mode, a
signal with a frequency of $f_{0} = 2.05$ GHz is injected from the
left port of the ST-MTM antenna, and a reference antenna located in the
far-field region of the ST-MTM antenna receives the fundamental and
harmonic frequencies in different angles. Fig. \ref{fig:Measurement}B and \ref{fig:Measurement}C show the normalized measured radiation patterns for sequence
111100000 when the signal is injected from the left and right port, respectively. Based on the aforementioned results shown in Fig. \ref{fig:Stransmit}, it is expected the main beam directions of fundamental and harmonic frequencies for both left- and right-port excitation should be the same, as can be observed in Fig. \ref{fig:Measurement}B and \ref{fig:Measurement}C. To verify the harmonic beam scanning, in Fig. \ref{fig:Measurement}D we change the sequence to 111111100, whereas in Fig. \ref{fig:Measurement}E the sequence is 110000000 with the signal injected from the left port. As can be seen clearly,  with different sequences, the main beam directions of the fundamental and harmonic tones can be steered. 

Moreover, Fig. \ref{fig:Measurement}F and \ref{fig:Measurement}G depict the measured patterns in the receive mode with the sequence 111100000, when the signal with a frequency of \(f_{0}\) is illuminated from a reference horn antenna to the ST-MTM antenna at different angles, in which the received spectral components are  collected from the left and right port, respectively. The received signals are then normalized to the maximum of the fundamental frequency. It is also expected that the harmonic patterns for the  transmit mode and receive mode should be the same, when excited at $f_{0}$, under the condition \(\mathrm{\Delta}f \ll f_{0}\ \), which can be clearly verified from the measured patterns. Furthermore, we then change the
illuminated signal frequency to \(2.05\ GHz +\) \(\mathrm{\Delta}f\),
where \(\mathrm{\Delta}f\) is $1.58$ MHz. The received patterns from the right port are shown in Fig. \ref{fig:Measurement}H, where it can be seen that the main beam of \(2.05\ GHz +\) \(\mathrm{\Delta}f\) is now at the broadside.

It is noticed that the simulated results are based on the antenna with
$12$ unit cells. Nevertheless, due to the fabrication limitations, we
demonstrate a ST-MTM antenna with 9 cells instead. A larger number of
MTM unit cells will result in increasing the directivity of the antenna with narrower main beams as can be observed in Fig. \ref{fig:HScanning}. While the simulation and measurement results agree with each other, there are some discrepancies in the main beam direction of the fundamental
and harmonic frequency patterns, mainly resulting from the difference in unit cell numbers as well as the propagation constant values of the two bias states between the measurement and the full-wave EM simulation. It is noted a small change in the bias voltages come from the FPGA may lead to some deviation of the propagation constant in the two states (0 and 1). 

%Moreover, although the measurement is conducted inside a microwave anechoic chamber, only a small reflection in the chamber can lead to changing the main beam direction since the beams are too wide due to the low number of cells.

According to the measured results shown in Fig. \ref{fig:Measurement}, simultaneous transmit and receive is
possible based on the proposed approach illustrated in Fig. \ref{fig:Stransmit}A, where \(\mathrm{\Delta}f \ll f_{0}\) for our proposed
ST-MTM antenna. On the other hand, the modulation frequency cannot be
more than $5.5$ MHz due to the hardware limitation of FPGA in the measurement, i.e. the cases when \(\mathrm{\Delta}f/f_{0} > 0.1.\)

\section*{Materials and Methods}
%\textcolor{red}{
For the theoretical results, MATLAB is used to plot the far field pattern of the fundamental and harmonic frequencies based on the derived equations. For the programmable unit cell design, HFSS as a full wave simulator, is utilized to design the MTM unit cell with dispersion diagrams shown in Supplementary, Fig. \ref{fig:SI1}B. The length of the unit cell is around $1.5$ cm. In the simulation, the varactors are modeled using a series $RLC$ circuit, where $R$=4.8\(\ \Omega\), $L$=$0.7$ nH, and $C$ varies with respect to the bias voltages. For the Bias ``$0$'' state the capacitor value is 1.05 pF and for Bias ``$1$'' state it is $2.1$ pF. The prototype is simulated and fabricated on a RO$5870$ substrate with a dielectric constant of $2.2$ and thickness of $1.57$ millimeter. The varactors are SMV2019 from Skyworks, where the capacitance range varies from $2.2$ pF at a bias voltage of $0$ V to $0.3$ pF at $20$V. It is noted $0.1$V and $3$V can provide our design with the required positive and negative phase constants, respectively. 

Basys $3$ FPGA board is used to provide $9$ digital outputs for the  control bias voltages of each unit cell. The low-level output voltage is around $50$ mV, whereas the high-level one is around $3.2$ V, which can provide us the 0 and 1 states for the predetermined sequences. The period of the time coding sequence is selected to be $0.63$ microsecond, resulting in a modulation frequency of $1.58$ MHz. The radiation pattern measurements are conducted inside a microwave anechoic chamber. A reference horn antenna is used as a transmitter when the prototype operates in the receive mode, and as a receiver when the prototype operates as a transmitter. A signal generator is utilized to provide a signal in the frequency of around $2$ GHz. By using a spectrum analyzer, the received signals at the fundamental and harmonic frequencies are collected from the reference horn antenna in the transmit mode, as well as from the left and right port of the prototype in the receive mode when illuminated from different angles.
%}
\section*{Discussion}

In this paper, we propose a digitally space-time-coded MTM antenna exhibiting harmonic beam scanning and nonreciprocity behavior,
which can be used for simultaneous transmit and receive. The phase constant of each varactor-embedded programmable CRLH unit cell changes between positive (state $1$), and negative (state $0$) values based on a
predetermined sequence in a periodic fashion in both space and time domain. In so doing, harmonic beam scanning can be realized by simply changing the programming coding sequences, which may lead to various applications including MIMO communication and radar detection.

%\textcolor{red}{(need to rewrite:) 
Moreover, nonreciprocal transmission and reception of harmonic waves can be achieved by the proposed ST-MTM antenna. When \(\mathrm{\Delta}f \ll f_{0}\ \), the generated harmonic signal at \(f_{0} + \mathrm{\Delta}f\) will radiate towards a specific direction in the transmit mode, while in the receive mode, by illuminating a signal at \(f_{0} + \mathrm{\Delta}f\) to the ST-MTM antenna at the same direction, the dominant generated harmonic signal becomes \(f_{0} + 2\mathrm{\Delta}f\). Leveraging such nonreciprocal property, simultaneous transmit and receive can therefore be realized by transmitting the signal at \(f_{0}\), whereas the information of the receiving signal can be extracted at the frequency of \(f_{0} + 2\mathrm{\Delta}f\).

%for the simultaneous transmit and receive operation, in the transmit mode a signal is injected in \(f_{0}\) frequency from the left port, generated harmonics are propagated by the ST-MTM antenna, a receiver located in the main beam of \(f_{0} + \mathrm{\Delta}f\) can receive the information in this frequency. In receive mode, \(f_{0} + \mathrm{\Delta}f\ \) is illuminated to the ST-MTM antenna, and harmonics are generated while signal is travelling inside the structure and propagate towards the right and left ports. 

%Finally, information can be extracted from the left or right port in \(f_{0} + 2\mathrm{\Delta}f\) frequency. 
On the other hand, when the modulation frequency is not much smaller than the fundamental frequency, different main beam directions of harmonic waves for transmit and receive patterns can be observed when injecting or illuminating the antenna at the same frequency. A prototype with 9 programmable MTM unit cells is fabricated, in which the experimental results validate the proposed concept of ST-MTM antenna.

%the fundamental frequency is injected from the left port and the main beam direction of transmitted harmonic is different than the direction that maximum power is received at that harmonic frequency from the left port in the receive mode.

\vspace{-1em}

\clearpage


\section*{Acknowledgements}

This work was supported by the National Science Foundation (NSF) under Grant ECCS-2229384
and ECCS-2028823. Any opinions, findings, and conclusions or recommendations expressed in this material are those of the authors and do not necessarily reflect the views of the National Science Foundation. 

\section*{Author contributions statement}

Conceptualization: C.-T.M.W.; design, prototyping, experiments and data collection: S.V; Writing—original draft: S.V.; writing—review \& editing: C.-T.M.W.

\section*{Data availability}
The data that support the findings of this study are available from the corresponding author upon reasonable request.

\section*{Additional information}
\textbf{Competing interests}: The authors declare no competing interests.

\newpage
\setcounter{figure}{0}
\renewcommand{\figurename}{Fig.}
\renewcommand{\thefigure}{S\arabic{figure}}
\hspace{-20pt}
Supplementary Information for
\\
\\
{\huge Programming nonreciprocity and harmonic beam steering via a digitally space-time-coded metamaterial antenna}
\\
\\

\vspace{8em}
\hspace{-15pt}\textbf{Far-field radiation pattern derivation in the transmit mode.}
For
expansion of $U_{n}(t)$ in the form of Fourier series, we start from
$H_{u}(t)$:

\begin{align*}
 H_{u}(t) = \sum_{q = - \infty}^{\infty}{a_{u}^{q}e^{j2\pi q\mathrm{\Delta}ft}}\ \ \ \ \ \ \ \ \ \ \ \ \ \ (1)   
\end{align*}

\begin{align*}
 U_{n}(t) = \sum_{u = 1}^{L}{\gamma_{n}^{u}H_{u}(t)} = \sum_{u = 1}^{L}\gamma_{n}^{u}\sum_{q = - \infty}^{\infty}{a_{u}^{q}e^{j2\pi q\mathrm{\Delta}ft}} = \sum_{q = - \infty}^{\infty}{\sum_{u = 1}^{L}\gamma_{n}^{u}a_{u}^{q}e^{j2\pi q\mathrm{\Delta}ft}} = \sum_{q = - \infty}^{\infty}{b_{u}^{q}e^{j2\pi q\mathrm{\Delta}ft}}\ \ \ \ (2)  
\end{align*}

So, $b_{u}^{q}$ is the Fourier series coefficient of the periodic
function $U_{n}(t)$ and is obtained as follows:

\begin{align*}
 b_{u}^{q} &= \sum_{u = 1}^{L}\gamma_{n}^{u}a_{u}^{q} = \sum_{u = 1}^{L}{\frac{\gamma_{n}^{u}}{T_{0}}\int_{\frac{(u - 1)T}{L}}^{\frac{uT}{L}}e^{- j2\pi q\mathrm{\Delta}ft}dt} = \sum_{u = 1}^{L}{\gamma_{n}^{u}f_{0}}\frac{e^{- j2\pi q\mathrm{\Delta}ft}}{- j2\pi q\mathrm{\Delta}f}\Biggr|_{\frac{(u - 1)T_{0}}{L}}^{\frac{uT_{0}}{L}} \
\\
&=\sum_{u = 1}^{L}\frac{\gamma_{n}^{u}}{- 2j\pi q}\left\lbrack e^{\frac{- j2\pi qu}{L}} - e^{\frac{- j2\pi q(u - 1)}{L}} \right\rbrack = \sum_{u = 1}^{L}\frac{\gamma_{n}^{u}}{- 2j\pi q}e^{\frac{- j\pi q(2u - 1)}{L}}\sin\left( \frac{\pi q}{L} \right)*( - 2j) \
\\
&=\sum_{u = 1}^{L}\frac{\gamma_{n}^{u}}{L}e^{\frac{- j\pi q(2u - 1)}{L}}{sinc}\left( \frac{\pi q}{L} \right) = e^{\frac{j\pi q}{L}}{sinc}{\left( \frac{\pi q}{L} \right)\sum_{u = 1}^{L}{e^{\frac{- j2\pi qu}{L}}\sum_{n = 1}^{N}\frac{\gamma_{n}^{u}}{L}}}\ \ \ \ \ \ \ \ \ \ \ \  
 (3)   
\end{align*}

\vspace{1em}
By substituting this expression in the far field radiation pattern, we
have:

\begin{align*}
R(\theta,t) &= S(t)\sum_{n = 1}^{N}{I_{0}e^{- \alpha(n - 1)p}e}^{jk_{q}(n - 1)pcos\theta}U_{n}(t) \\
&=e^{j2\pi f_{0}t}\sum_{q = - \infty}^{\infty}{e^{j2\pi qf_{0}t}{I_{0}e^{- \alpha(n - 1)p}e}^{jk_{q}(n - 1)pcos\theta}sinc\left( \frac{\pi q}{L} \right)e^{\frac{j\pi q}{L}}\sum_{u = 1}^{L}{e^{\frac{- j2\pi qu}{L}}\sum_{n = 1}^{N}\frac{\gamma_{n}^{u}}{L}}} \\
&=\sum_{q = - \infty}^{\infty}{e^{- j2\pi t(f_{0} - q\mathrm{\Delta}f)}sinc\left( \frac{\pi q}{L} \right)e^{\frac{j\pi q}{L}}\sum_{u = 1}^{L}{e^{\frac{- j2\pi qu}{L}}\sum_{n = 1}^{N}\frac{\gamma_{n}^{u}}{L}}}{I_{0}e^{- \alpha(n - 1)p}e}^{jk_{q}(n - 1)pcos\theta}\ \ \ (4)   
\end{align*}

\vspace{1em}
Therefore, equation (4) is obtained which is equation \eqref{eq:R1} in the
paper.

\vspace{2em}
\hspace{-15pt}\textbf{Formulation of patterns in the receive mode.}
In the receive mode, a
signal with frequency $f_{0}$ is illuminated to the ST-MTM antenna. The
fundamental and harmonic patterns are received from the left port.

To illustrate, the signal radiated from the $n^{th}\ $unit cell in the
transmit mode can be written as:
\begin{align*}
 R_{n}^{TX}(\theta,t) = \ S(t){I_{0}e^{- \alpha(n - 1)p}e}^{jk_{q}(n - 1)pcos\theta}U_{n}(t),\ \ {U_{n}(t) = e}^{j\sum_{m = 1}^{n}\varphi_{m(t)}} = \ \sum_{u = 1}^{L}{\prod_{m = 1}^{n}e^{- j\varphi_{m}^{u}}H_{u}(t)}\ \ (5)   
\end{align*}

In this case, the signal is injected from the left port and reaches to
the $n^{th}$ unit cell after passing from the previous ($n - 1)$ cells
which leads to $j\sum_{m = 1}^{n}\varphi_{m(t)}$ phase shift. Then some
portion of it radiates to the free space.

On the other hand, in the receive mode after capturing the signal
through $nth$ unit cell it travels toward left and right ports. The
portion of it that travels toward the left port, passes from the same
($n - 1)$ cells and reaches to the left port. In this case, the phase
shift will be $j\sum_{m = 1}^{n}\varphi_{m(t)}$ which is the same as
phase shift in transmit mode. As such, the signal received by the
$n^{th}$ unit cell which then reaches to the left port can be written
as:
\begin{align*}
  R_{n}^{RX}(\theta,t) = \ S(t){I_{0}e^{- \alpha(n - 1)p}e}^{jk_{0}(n - 1)pcos\theta}U_{n}(t),\ \ {U_{n}(t) = e}^{j\sum_{m = 1}^{n}\varphi_{m(t)}} = \ \sum_{u = 1}^{L}{\prod_{m = 1}^{n}e^{- j\varphi_{m}^{u}}H_{u}(t)}\ \ (6)
\end{align*}

The only difference of equation (5) with (6) is replacing $k_{q}$
with $k_{0} = 2\pi f_{0}/\ C$.
\clearpage
\begin{figure*}[h]
\centering
\includegraphics[width=12.5cm,height=7.8cm]{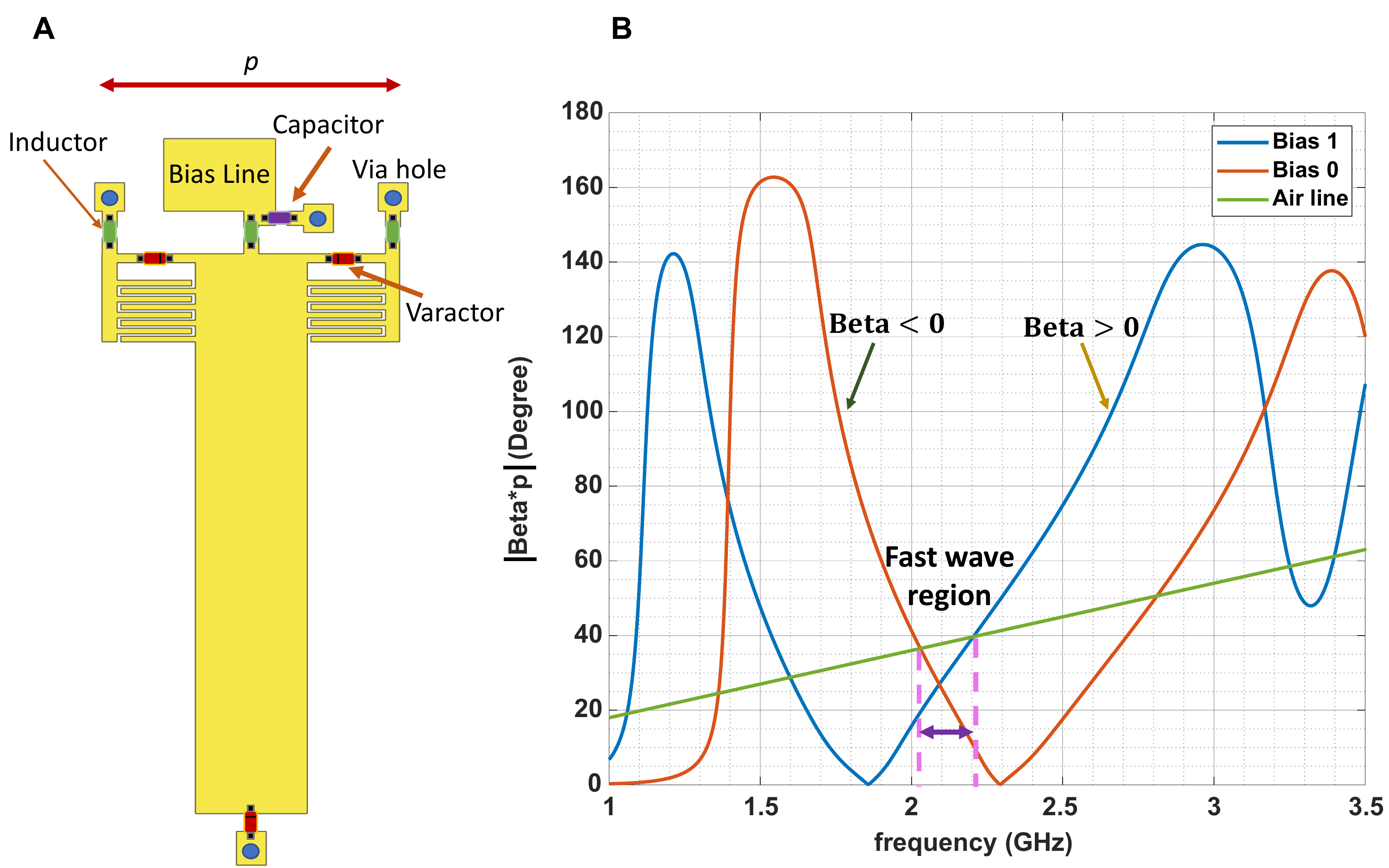}
\caption{
(A)Tunable unit cell schematic. (B) Simulated Dispersion
curves for two states of varactor Bias.}
\label{fig:SI1}
\end{figure*}

\begin{figure*}[h]
\vspace{2em}
\centering
\includegraphics[width=13.3cm,height=7cm]{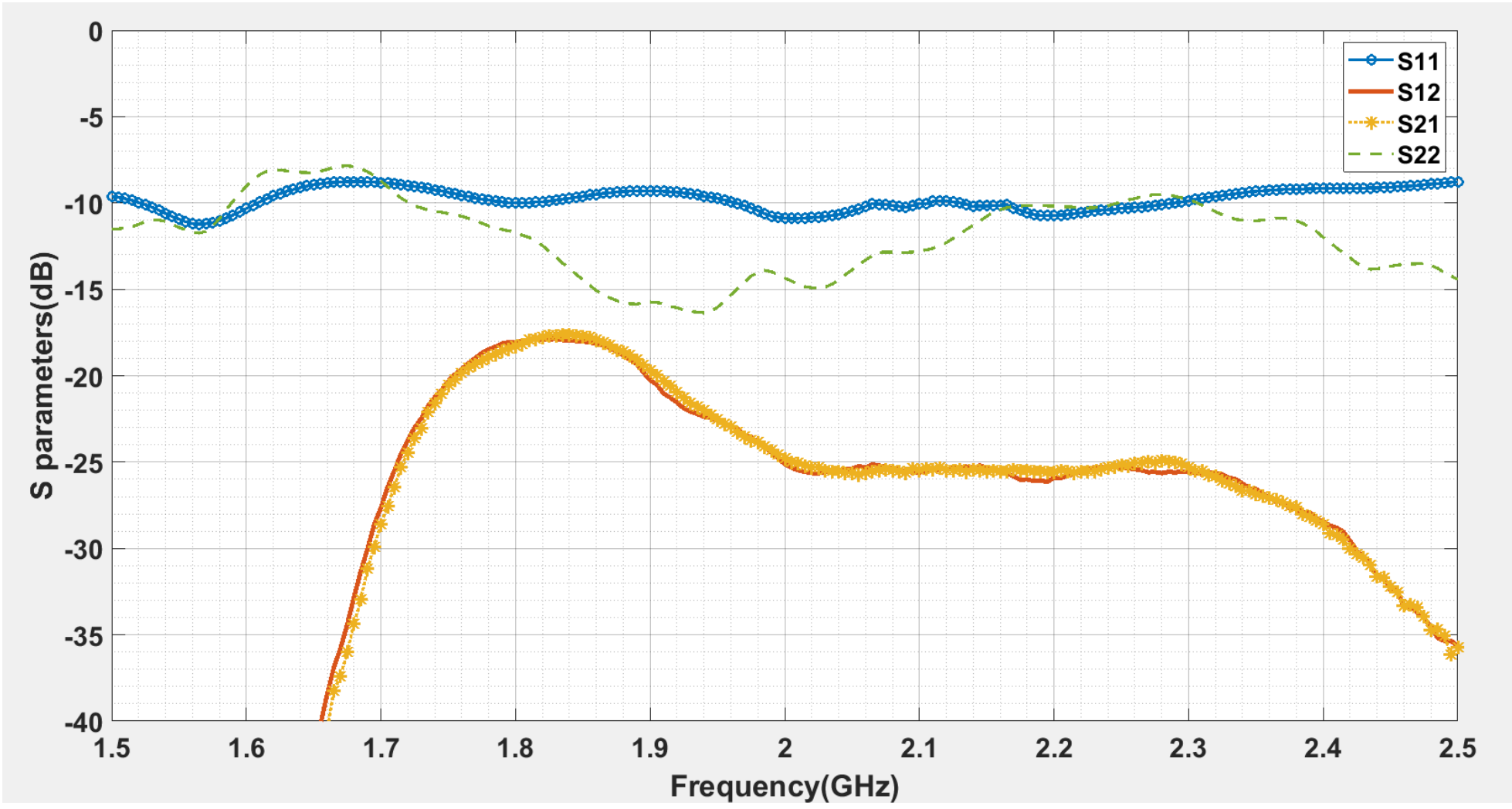}
\caption{
Measured S parameters of the fabricated TMLWA with 9 cells when the sequence in first time slot is 111100000 and then it is circulated one bit backward for the next time slots.} 
\label{fig:SI2}
\end{figure*}

\end{document}